\documentclass[12pt]{revtex4}
\usepackage{rotating}
\usepackage{graphicx}
\usepackage[english]{babel}
\usepackage{times}
\usepackage{mathptm}
\usepackage{supertabular}
\usepackage{dcolumn}
\usepackage{fancyhdr}

\renewcommand{\baselinestretch}{1.5}
\begin{document}
\renewcommand{\baselinestretch}{1.5}
\begin{titlepage}
\title{DIFFERENTIAL EQUATION  MODEL FOR THE  FIRST $0_1 ^+ \rightarrow 2_1 ^+ $ STATE EXCITATION 
ENERGY E2 OF EVEN-EVEN NUCLEI$*$ }

\author{ {      R. C. Nayak  } \\
{Department of Physics, Berhampur University, Brahmapur-760007,India.} \\ 
{ S. Pattnaik} \\
{ Taratarini  College, Purusottampur, Ganjam, Odisha, India.     }}


\renewcommand{\baselinestretch}{1.5}
\begin{abstract}
  We propose here a new model termed as the Differential Equation Model
 for the first $0_1 ^+ \rightarrow 2_1 ^+ $ state  excitation energy E2 of 
a given even-even nucleus,  according to which   the energy E2  is expressed in terms of its
derivatives with respect to the neutron and proton numbers. This is based on 
a similar derivative equation satisfied by its complementary physical
quantity namely the Reduced
Electric Quadrupole  Transition Probability B(E2)$\uparrow$ in a recently 
developed model.
 Although the proposed differential equation for E2 
has been  perceived  on the basis of its close similarity to
B(E2)$\uparrow$,
 its theoretical foundation otherwise has been clearly demonstrated.
 We further exploit  
the very definitions of the  derivatives occurring in the 
 differential equation  in the model
 to obtain two different recursion relations for    E2,
connecting  in each case  three neighboring even-even nuclei from lower to
 higher mass 
numbers and vice-verse. We  demonstrate their numerical validity using the
available data throughout the nuclear chart  and also explore their
possible  utility in predicting the unknown E2 values.

\end{abstract}
\maketitle
\vskip 5cm
$*$ This is a slightly modified version of the article submitted for publication in Int.  Jou. of Mod. Phys. (2014).

\end{titlepage}
\newpage
\section{Introduction}

Reduced electric quadrupole transition probability B(E2)$\uparrow$ and its 
complimentary quantity, namely  the first $0_1 ^+ \rightarrow 2_1 ^+ $ state  excitation energy E2
for a even-even nucleus    play 
crucial  roles  for the study of excited states of nuclei and more
importantly the inherent nuclear structure.
Such studies got  a boost with the  
 advent of isotope facilities providing a large amount of  experimental 
data for several nuclides throughout the nuclear chart.
 The existence of a large  volume  of experimental data  led Raman et al.
 \cite{rmn} at the Oak Ridge 
Nuclear Data Project \cite{rmn,rm3} to make a comprehensive
analysis of all those data, in  preparing  the most sought-after experimentally
 adopted data table for both the above two physical quantities.
Of late,  
Pritychenko et al. \cite{prt} followed the process in compiling the newly emerging  data
sets  for  even-even
nuclei  near  $ N \sim Z \sim $ 28. These new data including the old set obviously put a 
challenge for the nuclear theorists to understand them.

Theoretically, possible existence of 
symmetry in 
 nuclear dynamics first explored in developing  mass formulas such as the 
Garvey-Kelson \cite{gkl} mass
formula that connects masses of six neighboring nuclei, got into the domain of
properties of the
 excited states. In this regard possible existence of such  symmetry 
 led Ross and Bhaduri\cite{rbh} in developing difference
equations involving both B(E2)$\uparrow$ and the E2 excitation energies 
 of the neighboring even-even nuclei.
Patnaik et al. \cite{pat}
 on the other hand have  also succeeded   in establishing
even more  simpler difference equations connecting  these values of
four neighboring even-even nuclei. 
 
 Just recently  we\cite{dem} have succeeded in  developing a new model for
 the B(E2)$\uparrow$,
termed as the Differential Equation Model (DEM) according to which,
the B(E2)$\uparrow$  value of 
 a given even-even nucleus is expressed in terms of  its derivatives 
with respect to the neutron and proton numbers. Since these two quantities 
more or less complement each other, it is expected that the  excitation 
energy E2 should also satisfy a similar differential equation.
Therefore in the present work with this view in the background, we propose
a similar model  for  E2 and explore its validity
and utility in predicting hitherto unknown  data.
It is needless to stress  here that
 any relation in the form of a differential equation of any physical
quantity is intrinsically sound enough to posses a good predictive ability. 
This
philosophy has been well demonstrated   in case of B(E2)$\uparrow$ 
predictions \cite{dem} just recently,   and also over the recent years in the 
development \cite{inm,in2,in3,in5} of  
the Infinite Nuclear Matter (INM) model of atomic nuclei  specifically for
the prediction \cite{in5} of nuclear masses. We should also note here that the
development of the Differential Equation Model for the B(E2)$\uparrow$ and
presently for E2 is also
based on the local energy relation of the INM model, which happens to be 
 an important component of the ground-state energy of a nucleus signifying
its individual characteristic nature.

   In Sec.
 II, we show how such a   relation in the form of a differential equation for 
E2 can be formulated 
 followed by its possible theoretical justification.  Sec. III deals
with how the same  differential  equation can be used to derive two
recursion relations in E2, connecting in each case three different
 neighboring even-even nuclei.
 Finally we present in 
 Section IV, their  numerical validity   when subjected to the
known \cite{rmn} experimental data throughout the nuclear chart, and their 
possible utility in   predicting its unknown values.

\section{Derivation of the Differential Equation for   the First $0_1^+ \rightarrow 2_1^+$ 
State Excitation Energy E2 }

As mentioned above that the development of the DEM model both for the 
B(E2)$\uparrow$ and presently for E2 owes its origin to the local energy 
differential equation in the INM model of atomic nuclei.
Physically the local energy $\eta$ embodies all the characteristic 
properties of a given nucleus, mainly the shell and deformation, and has been
explicitly shown \cite{rcn} to carry the shell-structure. Therefore  it is 
 likely  to have some characteristic correspondence   with  the 
properties  of excited states of a given
nucleus in general and in particular,  the reduced transition probability 
$B(E2)\uparrow$ and its complementary quantity E2.
Accordingly 
 the  $\eta$-equation as well as the B(E2)$\uparrow$-equation [see for instance
 the Eqs. (1 and 4) of the DEM model \cite{dem}] can be used as an ansatz to 
satisfy a
similar relation involving the E2 value of a given even-even nucleus. As a 
result  we write on analogy, 
a similar  equation for E2 as
\begin{equation}
  \label{be2} {E2[N,Z]/ A}={1\over 2} \left[(1+\beta){\Bigl(\partial E2
  / \partial N\Bigr)}_Z + (1-\beta){\Bigl(\partial E2/  \partial Z}\Bigr)_N\right].
\end{equation}
Thus we see that we have a relation (\ref{be2}) that  connects the E2
 value of a given nucleus (N,Z)
with its partial derivatives with respect to neutron   and proton 
numbers N and  Z. It is true that our   proposition of  this differential
 equation for E2 is purely   on the basis of 
intuition and on close analogy with that of B(E2)$\uparrow$.
  However  its  validity needs to be 
established, which  we show in the following. 

\begin{figure}[bth]
\includegraphics[width=6.3in, height=7.0in,angle=-0]{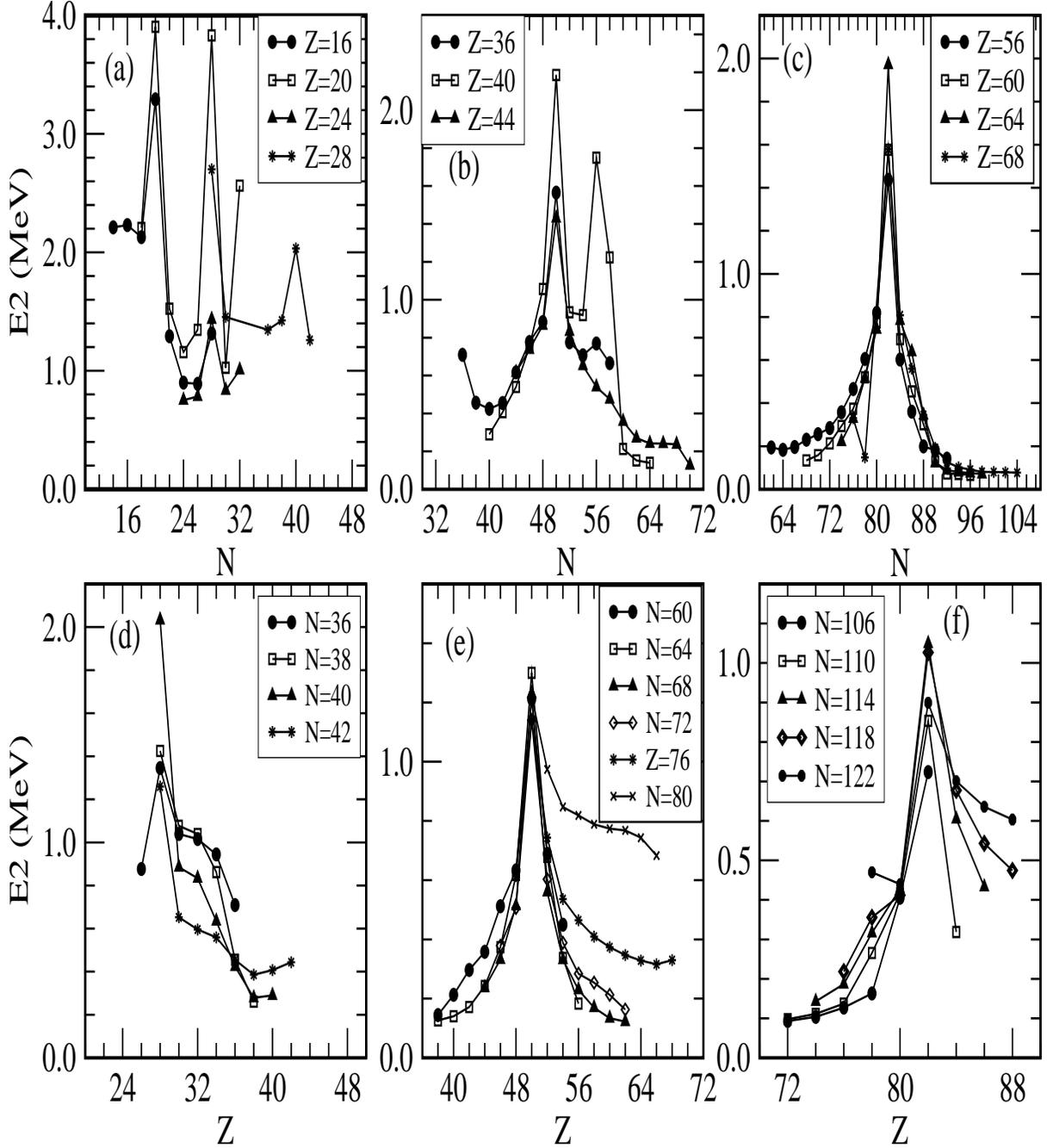}
\renewcommand{\baselinestretch}{1.}
\caption{Known E2 values  plotted as isolines for even-even nuclei.
 Isolines drawn  in the graphs (a-c)  connect E2  values
of various  isotopes for Z=16 on wards with varying neutron number N,
 while the isolines drawn  in the graphs (d-f) show the same for 
isotones for N=36 on wards with varying proton number Z. Other possible  
isolines are not
 shown here  to avoid clumsiness of the graphs.}
\end{figure}

For a theoretical justification of the above equation, we use the
 approximation  of expressing  E2  as the sum of two 
different functions $ E_1(N)$ and  $ E_2(Z)$ as
\begin{equation}
  \label{f} {E2[N,Z]}= E_1(N) + E_2(Z).
\end{equation}
The goodness of this simplistic approximation can only be judged from numerical
analysis of the resulting equations that follow using the experimental data. 
 Secondly we use the empirical fact [see Fig. 1] that  E2s are
 more or less slowly varying functions of N 
and Z  locally. This assumption however cannot be strictly true
  at the magic 
numbers and in regions where  deformations drastically change. In fact known
  E2 values plotted as isolines for  isotopes and isotones 
 in Fig. 1, convincingly demonstrate the above  aspects in most of the 
cases.  The usual typical bending  and kinks at magic numbers like 20, 50, 82
and semi-magic numbers 28 and 40  can be seen as a result of  sharply 
changing deformations.  Consequently
 $E_1$ and $E_2$  can be written directly
proportional to N and Z respectively as 
\begin{equation}
\label{f2}
  E_1(N)= \lambda N  ,
~~and~~   E_2(Z) =\nu Z,
\end{equation}
where $\lambda$  and $\nu$ are  arbitrary constants and vary from branch to
branch across the kinks. Then one can
easily see that  just by substitution of the above two  Eqs. (\ref{f},\ref{f2}),
 the differential Eq. (\ref{be2}) gets directly satisfied.
 Thus  the proposed differential  equation for
E2 analogous to the  B(E2)$\uparrow$
relation in the DEM model gets theoretically justified. 
 However, the  differential Eq.
 (\ref{be2}) has its own limitations, and need  not be expected to remain 
strictly valid across the   magic-number nuclei
 because of the very approximations involved in proving it.

\section{Derivation of the Recursion Relations in    E2}

It is always desirable  to solve  the differential  Eq. (\ref{be2}) in order  to utilize it  for  practical applications.
Therefore it is necessary  to obtain possible  recursion relations in  E2 
for even-even nuclei in (N,Z) space from it.
The partial derivatives occurring in this equation  at mathematical
 level are defined for continuous
 functions. However for finite nuclei, these derivatives  are to be
evaluated taking the difference of E2 values
of neighboring nuclei. Since our interest is to obtain recursion relations for
even-even nuclei,
  we use in the above equation  the usual 
forward and backward definitions for the  partial  derivatives. These are
   given by
\begin{eqnarray}
  \label{der} 
\Bigl({\partial E2/
  \partial N}\Bigr)_Z &\simeq&{1\over 2}  \Bigl[ E2[N+2,Z]-E2
  [N,Z]\Bigr], \nonumber \\ 
\Bigl( {\partial
  E2/ \partial Z}\Bigr)_N &\simeq&{1\over 2}  \Bigl[ E2
  [N,Z+2]-E2[N,Z] \Bigr] ,\\
& & and \nonumber \\
\Bigl({\partial E2/
  \partial N}\Bigr)_Z &\simeq&{1\over 2}  \Bigl[ E2[N,Z]-E2
  [N-2,Z]\Bigr], \nonumber \\ 
\Bigl( {\partial
  E2/ \partial Z}\Bigr)_N &\simeq&{1\over 2}  \Bigl[ E2
  [N,Z]-E2[N,Z-2] \Bigr] .
\end{eqnarray}
 Substitution of the above two pairs of   definitions for the derivatives in 
the differential equation  (\ref{be2})   enabled us to derive
 the following two recursion 
relations for E2, each  connecting three  neighboring even-even nuclei.
These are
\begin{eqnarray}
\label{b2f}
E2[N,Z] &=& {N \over {A-2}}\: E2 [N-2,Z] + {Z \over {A-2}}\:E2 
[N,Z-2]  ,\\
\label{b2b}
 E2 [N,Z] &=& {N \over {A+2}}\: E2 [N+2,Z]+{Z \over {A+2}}\: E2
[N,Z+2] ..
\end {eqnarray}
The first recursion  relation (\ref{b2f}) connects three neighboring nuclei (N,Z), (N-2,Z)
and (N,Z-2) while the second one (\ref{b2b}) connects (N,Z), (N,Z+2) and (N+2,Z). 
  The first one  relates  E2  
of lower  to higher mass nuclei while the second one 
relates higher to lower mass, and hence they can be termed as the forward and 
backward recursion relations  termed  as
 E2-F and E2-B 
respectively. Thus   depending on the availability of E2 data, one can use 
either or both  of these two relations to obtain the corresponding
unknown  values of   neighboring nuclei.

\begin{figure}[bth]
\includegraphics[width=5.in, height=5in,angle=-0]{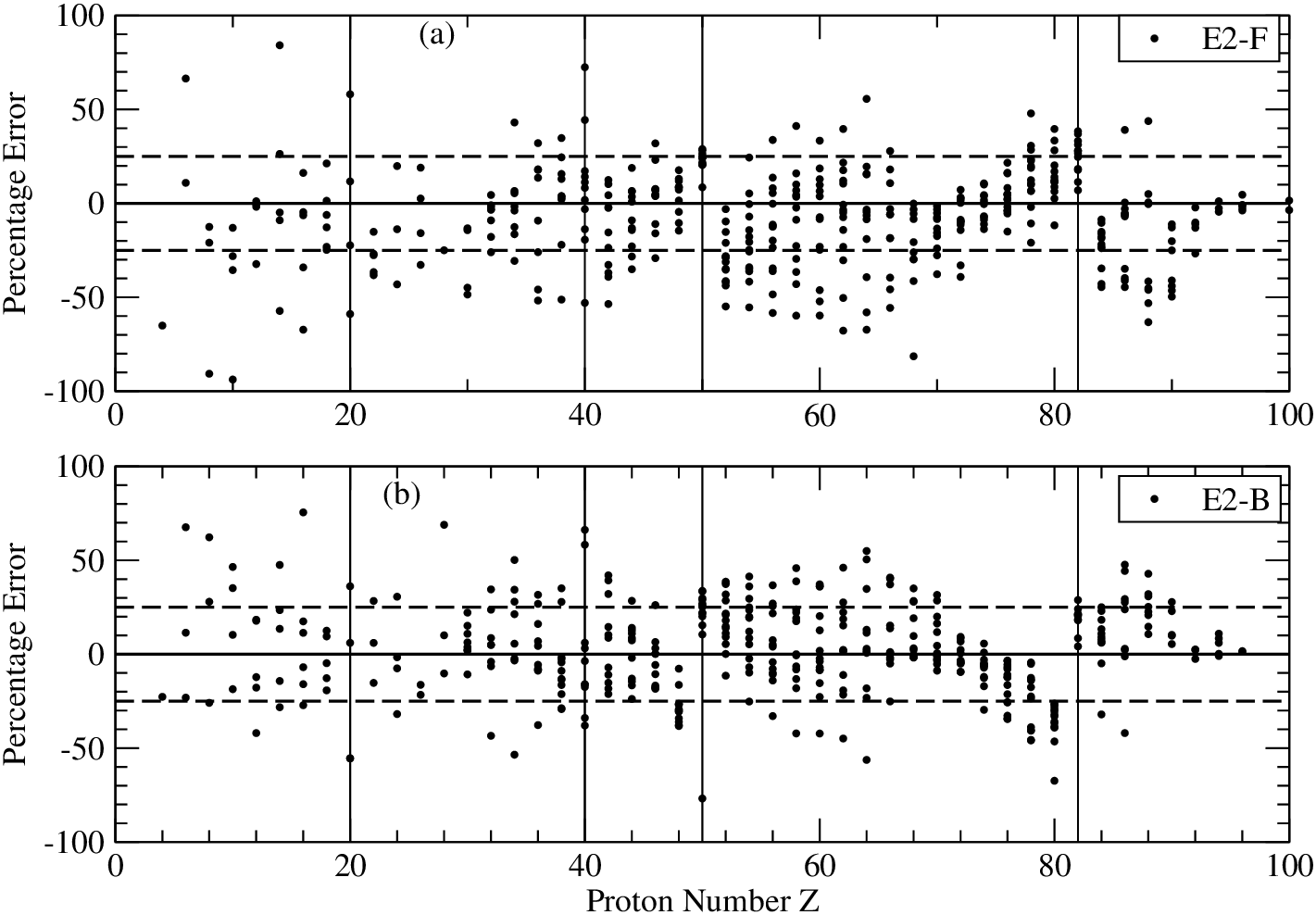}
\renewcommand{\baselinestretch}{1.}
\caption{Numerical Test of  the   Recursion Relations connecting E2 values 
of 
 neighboring nuclei. 
 The  percentage errors of the computed E2 values of all the
anchor nuclei  are plotted
 against  Proton  Number Z of those nuclei.
The graph (a)   shown as  E2-F 
corresponds to the results of the relation (\ref{b2f}) while graph (b)
marked as E2-B shows  those of the relation (\ref{b2b}). 
 The vertical 
solid lines are drawn   just  to focus  larger deviations if any
 at the magic and semi-magic numbers. }
\end{figure}

\section{ Numerical  Test of the  Recursion Relations in E2}

\begin{figure}[b!t!h!]
\includegraphics[width=5.in, height=5in,angle=-0]{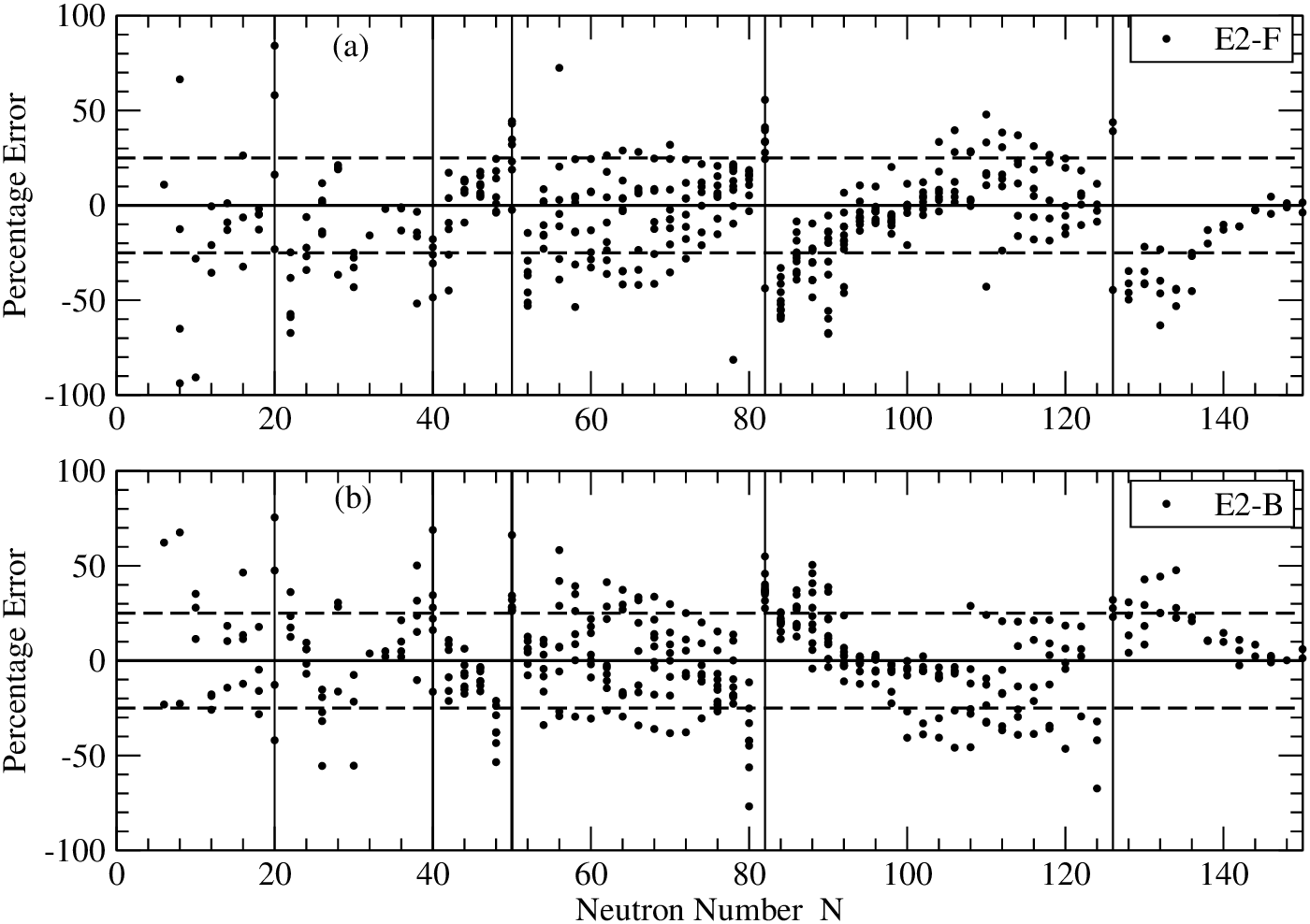}
\renewcommand{\baselinestretch}{1.}
\caption{Same as Fig.2 plotted versus Neutron Number N.}
\end{figure}
Having derived the recursion relations in E2 from the 
 differential equation (\ref{der}), it is desirable  to establish  their 
numerical validity to see to what extent they 
satisfy the known experimental data throughout the nuclear chart. This would
also numerically support the differential equation  (\ref{be2}) from which
 the recursion 
relations are derived.  For this purpose we use 
the experimentally adopted  E2 data set of   Raman et al. \cite{rmn} 
in the above  relations throughout  mass range of A=10 to 240
, and
compute the same  of all possible   anchor nuclei that   are
 characterized by the neutron and proton numbers (N,Z) occurring in the left
  hand sides of the  
relations (\ref{b2f},\ref{b2b}).
For better visualization of our results, we calculate  the deviations of the 
computed    E2 values   from  those of  the experimental  data 
     in terms of the   percentage errors 
  following Raman et al. \cite{rm2}. 
The percentage error of a particular calculated quantity is  as usual
defined as the  deviation of that quantity  from
  that of the experiment
   divided by the average of
 the concerned data inputs,  and then expressed as the percentage of the average.
Obviously the larger the percentage error  larger is the deviation of the 
concerned computed value.
  These percentages so computed are plotted
 in the figures  2 and
3 against the proton and neutron numbers respectively. This is intentionally 
done to ascertain
to what extent   possible  deviations occur   at proton and neutron magic
numbers.  
From the  presented results we see,  that in most of the cases both forward
and backward recursion relations  (\ref{b2f},\ref{b2b}) 
 give  reasonably good agreement with experiment.
 Numerically the deviations in  284  out of 417 cases for  
 the forward relation [E2-F]  and 278 out of 416  cases for the backward
relation  [E2-B] lie within 
$\pm$25\% error [shown   within broken lines in the figures]. 
  In view of this, the agreement of the model recursion relations with those
of experiment can be considered  good. 
 However one can see from the figures 2 and 3, that  the percentage errors (deviations) are relatively
higher for some nuclei in the neighborhood of the  magic numbers 20, 50, 82, 126
and semi-magic number 40.  Such increase in the vicinity of the magic 
numbers is  expected, as the differential Eq. (\ref{be2}) from which the 
recursion relations are derived need not be  strictly valid
 at the magic numbers.

\begin{figure}[b!t!h!]
\includegraphics[width=4in, height=5in,angle=-0]{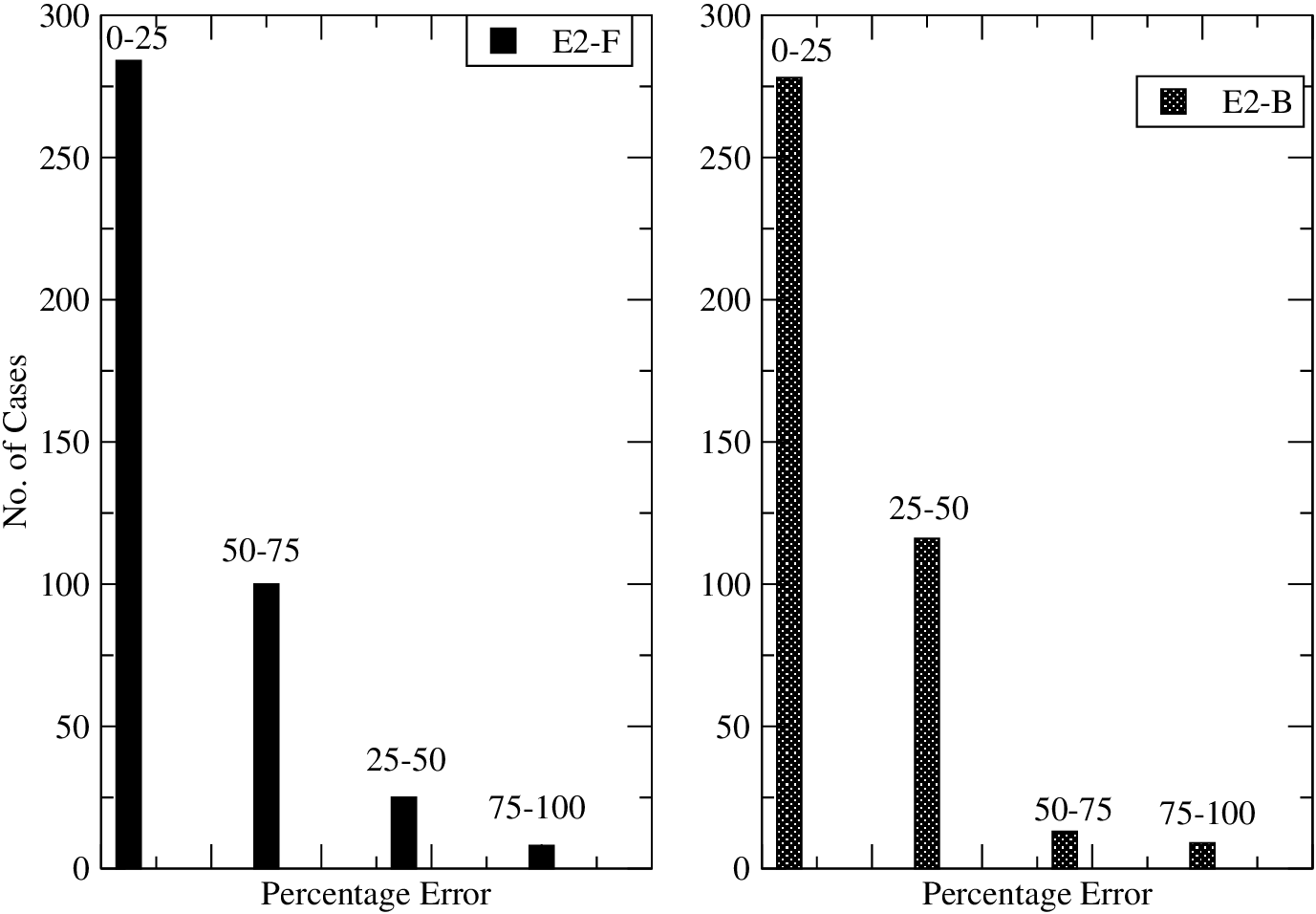}
\renewcommand{\baselinestretch}{1.}
\caption{Vertical pillars
 showing  number of cases having different ranges of absolute percentage 
 errors. Those  marked as E2-F and   E2-B correspond
to  the results of our relations (\ref{b2f} and \ref{b2b}) respectively.}
\end{figure}

 To  bring out the contrasting features of our results in a better way, 
   we also present our results   
 in the form of histograms in Fig. 4, which displays  the total  number of 
cases having
different ranges of percentage errors. As can be seen,  the sharply decreasing
  heights of the vertical pillars with the increasing  range of 
 errors  are a clear testimony of the goodness  of our  recursion 
relations.

For exact numerical comparison,
   we  also present in Table I  results  obtained in our calculation
 along with those of  the experiment \cite{rmn}
for some of the nuclides randomly chosen all over the  nuclear chart.
 One can easily see that the agreement of the predictions with the
 measured values is
rather   good in most of the cases. In few cases such as 
  $^{36}Si$, $^{40}S$, $^{44}Ca$ and $^{220}Rn$ there exists  little bit of 
 discrepancy in between the predictions of the relation (\ref{b2b}) and 
those of the experiment.
Thus  one can fairly say, that the overall agreement of our model predictions 
with those of experiment is exceedingly good.
Therefore  such agreement  is a clear testimony of the goodness  
of our recursion relations.


\begin{table}[h!t!b!]
\begin{center}
\begin{tabular}{|c c c c c c c c|}
\hline
\multicolumn{1}{|c } {Nucleus} &
\multicolumn{1}{ c }{Experiment} &
\multicolumn{1}{ c }{B(E2)-B} &
\multicolumn{1}{ c }{B(E2)-F} &
\multicolumn{1}{ c } {Nucleus} &
\multicolumn{1}{ c }{Experiment} &
\multicolumn{1}{ c }{B(E2)-B} &
\multicolumn{1}{ c|}{B(E2)-F} \\
\multicolumn{1}{|c } { } &
\multicolumn{1}{ c }{(MeV)} &
\multicolumn{1}{ c }{(MeV)} &
\multicolumn{1}{ c }{(MeV)} &
\multicolumn{1}{ c } { } &
\multicolumn{1}{ c }{(MeV)} &
\multicolumn{1}{ c }{(MeV)} &
\multicolumn{1}{ c|}{(MeV)} \\
\hline
$ ^{24}Ne$ &1.982 &- & 1.782 & $ ^{30}Mg$ &1.482  &1.509 & 1.226   \\
$ ^{36}Si$ &1.399   &2.429    &1.104   & $ ^{40}S$ &0.900 &1.272   &0.969  \\
$ ^{44}Ar$ &1.144 & 1.129  &1.403   & $ ^{44}Ca$ &1.157 & 1.447   & 1.089   \\
$ ^{48}Ti$ & 0.984  &1.146  &1.152   &  $ ^{74}Ge$ & 0.596 &0.776 &0.546 \\
$ ^{80}Se$ & 0.666  &0.632  &0.689 & $ ^{82}Kr$ & 0.776  &0.654 &0.823 \\
$ ^{92}Sr$ & 0.815   &0.798   &0.852 & $ ^{102}Zr $ &0.152    & 0.184 &0.158  \\
$ ^{106}Mo$ & 0.172   &0.175   &0.209 & $ ^{110}Ru$ & 0.241 &0.227  &0.277 \\
$ ^{114}Pd$ & 0.332  &0.309  &0.403 & $ ^{120}Cd$ &0.506 &0.452   &0.785  \\
$ ^{128}Sn$ &1.169   & 0.965  &1.056  & $ ^{130}Te$ &0.839    &0.928  &0.839 \\
$ ^{134}Xe$ &0.847 &0.803   &1.097     & $ ^{146}Ba$ &0.181  &0.223  &0.146  \\
$ ^{148}Ce$ &0.159   &0.231 &0.109 & $ ^{154}Nd$ &0.071 & 0.077 &0.070  \\
$ ^{158}Sm$ &0.073  & 0.073  &0.072     & $ ^{160}Gd$ &0.075 &0.078 & 0.074  \\
$ ^{164}Dy$ &0.073   &0.078 & 0.077 & $ ^{168}Er$ &0.080   &0.080  &0.080 \\
$ ^{174}Yb$ &0.077  &0.079  & 0.084  & $ ^{180}Hf$ &0.093  &0.091& 0.098 \\
$ ^{186}W$ &0.122  &0.111  & 0.146  & $ ^{192}Os$ &0.206  &0.196 & 0.259 \\
$ ^{196}Pt$ &0.356 &0.288  & 0.405  & $ ^{202}Hg$ &0.438&0.413  &0.614 \\
 $ ^{204}Pb$ &0.899  &0.759   &0.755 & $ ^{214}Po$ &0.609   &0.765   &0.511\\
 $ ^{220}Rn$ &0.241 &0.401  &0.155   & $ ^{224}Ra$ &0.084  &0.142    &0.069 \\
 $^{232}Th$ &0.049 &0.055  &0.047   & $^{236}U$ &0.045   & 0.046 & 0.044  \\
 $ ^{242}Pu$ &0.045  &0.044  &0.046 & $ ^{246}Cm$ &0.043  &0.045 &0.042  \\ 
 $ ^{250}Cf$ &0.043  &0.043  &0.046 & $ ^{254}Fm$ &0.045  &0.047  &-      \\ 
\hline
\end{tabular}
\caption{Comparison of the model predictions referred to as E2-B [Eq. (\ref{b2b})]
and  E2-F  [Eq. (\ref{b2f})] and the experimental\cite{rmn}  E2
 values.} 
\end{center}
\end{table}
 
Once we establish the goodness of the two recursion relations, it is desirable
 to compare our predictions with
 the latest experimentally adopted data  of  Pritychenko et al. \cite{prt}. 
  It must be made
clear that none of the values of the new experimental data set has been
 used in our recursion relations. Rather  we 
use only the available data set of Raman et al. \cite{rmn} to generate all possible values of a given
nucleus employing the two recursions relations 
 (\ref{b2f}) and (\ref{b2b}). One  should  note here that each of these
relations can be rewritten in three different ways just by shifting the three
terms occurring in  them from left to right and vice-verse. 
Thus altogether,  these two relations in principle can generate up to six
 alternate
values for a given nucleus  subject to availability of the corresponding data.
Since each of the values is equally probable, 
the predicted value for a given nucleus is then obtained by the arithmetic  
mean of all those  generated values so obtained. 
  Our predictions here are confined  to those
isotopes for which  measured values were quoted by Pritychenko et al. \cite{prt}.
The  predicted values so obtained  termed as  the 
DEM  values  are presented  in Table II
 for various isotopes of Z=24, 26, 28 and 30 
 along with those of the latest  experimental \cite{prt} data.

\begin{table}[h!t!b!]
\begin{center}
\begin{tabular}{|c c c c c c|}
\hline
\multicolumn{1}{|c}{ Nucleus} &
\multicolumn{1}{c}{Experiment \cite{prt}} &
\multicolumn{1}{c|}{DEM} &
\multicolumn{1}{c}{ Nucleus} &
\multicolumn{1}{c}{Experiment \cite{prt}} &
\multicolumn{1}{c|}{DEM} \\
\multicolumn{1}{|c } { } &
\multicolumn{1}{ c }{(MeV)} &
\multicolumn{1}{ c }{(MeV)} &
\multicolumn{1}{ c } { } &
\multicolumn{1}{ c }{(MeV)} &
\multicolumn{1}{ c|}{(MeV)} \\
\hline
$ ^{46}Cr$ & 0.892 &0.964 & $ ^{48}Cr$ & 0.752  &0.723   \\
$ ^{50}Cr$ & 0.783  &0.870  & $ ^{52}Cr$ & 1.434 &1.312   \\
$ ^{54}Cr$ & 0.835 &0.977   & $ ^{56}Cr$ & 1.007 &0.793    \\
$ ^{58}Cr$ & 0.881  &0.997  & $ ^{48}Fe$ & 0.970 & 0.805   \\
$ ^{50}Fe$ & 0.765 & 0.817 & $ ^{52}Fe$ & 0.850   & 0.752   \\
$ ^{54}Fe$ & 1.408 & 1.476 & $ ^{56}Fe$ & 0.847   & 0.645   \\
$ ^{58}Fe$ & 0.811 & 0.664 & $ ^{54}Ni$ &1.392  & 2.078   \\
$ ^{56}Ni$ &2.701  & 2.023 & $ ^{58}Ni$ &1.454  & 1.412  \\
$ ^{60}Ni$ &1.336  & 1.222 & $ ^{62}Ni$ &1.173  & 1.140  \\
$ ^{64}Ni$ &1.348  & 1.123& $ ^{66}Ni$ &1.425  & 1.431  \\
$ ^{68}Ni$ &2.034 & 1.133  & $ ^{70}Ni$ &1.260  & 1.611   \\
$ ^{72}Ni$ &1.096 & 1.111  & $ ^{74}Ni$ &1.024 & 0.548   \\
$ ^{76}Ni$ &0.992 & 0.890  & $ ^{62}Zn$ &0.954 & 0.918   \\
$ ^{64}Zn$ &0.992 &0.929   & $ ^{66}Zn$ &1.039 &1.065   \\
$ ^{68}Zn$ &1.077 &0.832  & $ ^{70}Zn$ &0.885 &1.2336   \\
$ ^{72}Zn$ &0.653  &0.717 & $ ^{74}Zn$ &0.606  &0.592  \\
$ ^{76}Zn$ &0.599 &0.664 & $ ^{78}Zn$ &0.730 &0.645  \\
$ ^{80}Zn$ &1.492 &1.572 &  &  &  \\
\hline
\end{tabular}
\end{center}
\renewcommand{\baselinestretch}{1.}
\caption{Comparison of the model predictions  and the latest experimental \cite{prt}
  E2 values.  } 
\end{table}

 We also present our DEM predictions   in  Fig. 5 to convey a  better 
visualization 
 of our  results. 
 One can easily see that in all  the cases except for 
 $ ^{54,68,74}Ni$ and  $^{68,70}Zn$, the agreement
 between the predictions
with those of the experiment are remarkably  good. For these few nuclei, the 
 discrepancies  may be attributed to the possible
 sub-shell effect as either proton or  neutron  numbers or both are close
to  semi-magic numbers 28 and 40. 
For sake of comparison we have also presented in Fig. 5, results obtained
 from two shell-model calculations \cite{gxp,prt} marked here as SM1 and SM2.
  One should note here, that the first one obtained
using the  effective interactions  GXPF1A \cite{gxp} 
 did  not succeed in getting reliable values for nuclei having 
neutron number beyond N=36 because of its own limitations.
 Hence the second shell-model with JUN45 effective 
interaction was performed by
 Pritychenko et al. \cite{prt} 
for the nuclei  $ ^{64}Fe, ^{68}Ni$ and  $^{72,74}Zn$. One can 
easily see that the shell-model values SM1 almost agree with those of ours
 for almost all the isotopes while the other shell-model values SM2  are 
rather away from ours as well as from the experimental values.

\begin{figure}[b!t!h!]
\includegraphics[width=5.in, height=6.0in,angle=-0]{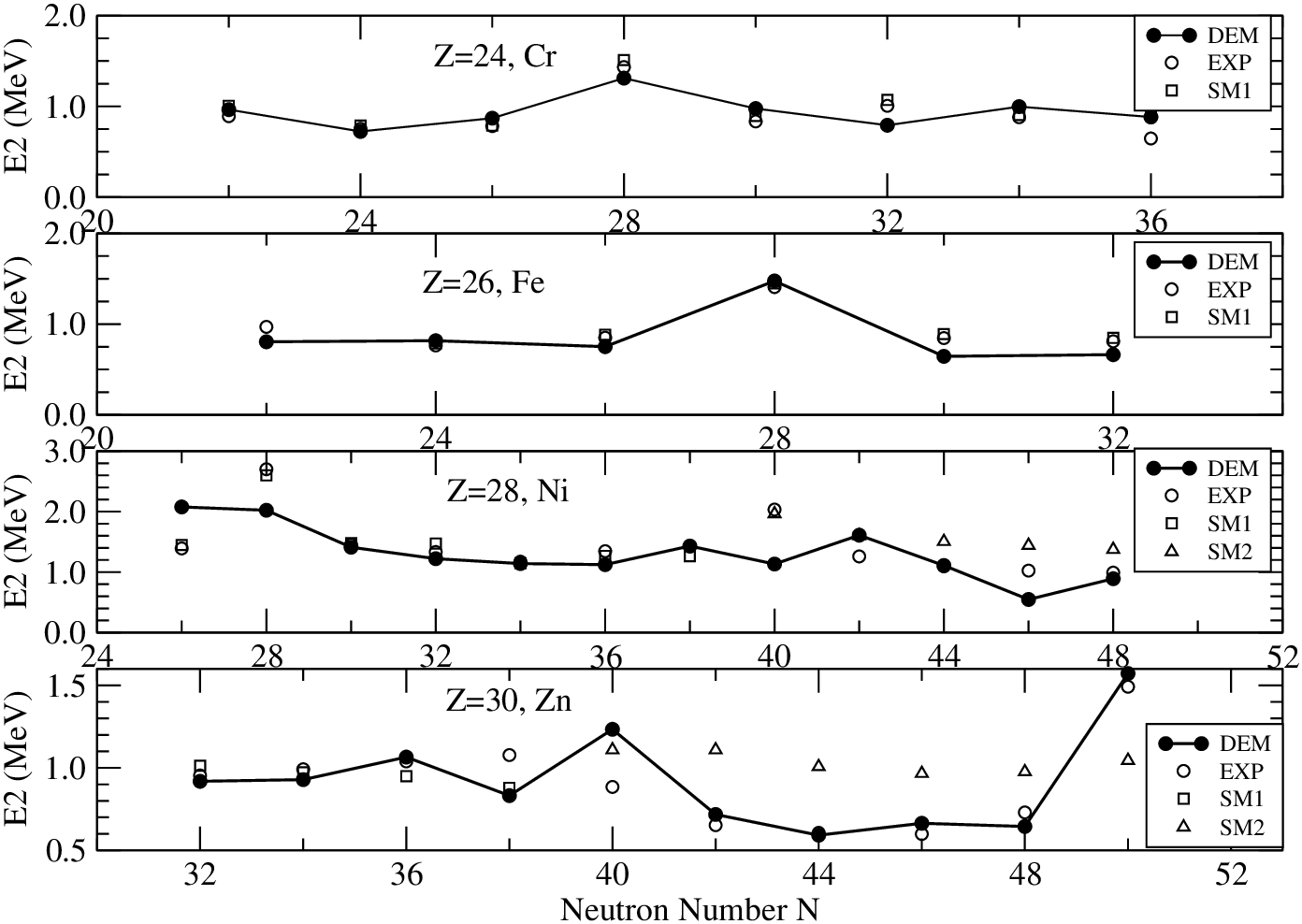}
\renewcommand{\baselinestretch}{1.}
\caption{Both calculated (DEM) and the latest experimental (EXP) \cite{prt} 
E2 values 
 [see text for details]  
are presented   for various isotopes of Z=24, 26,
 28 and 30  versus 
 Neutron Numbers N. DEM Values for different isotopes  are connected  by solid 
  lines just to guide the eye. 
 Recent shell-model calculated values [SM1 and SM2]  are also presented
for sake of comparison}
\end{figure}
	
\begin{figure}[bth]
\includegraphics[width=6.in, height=7.0in,angle=-0]{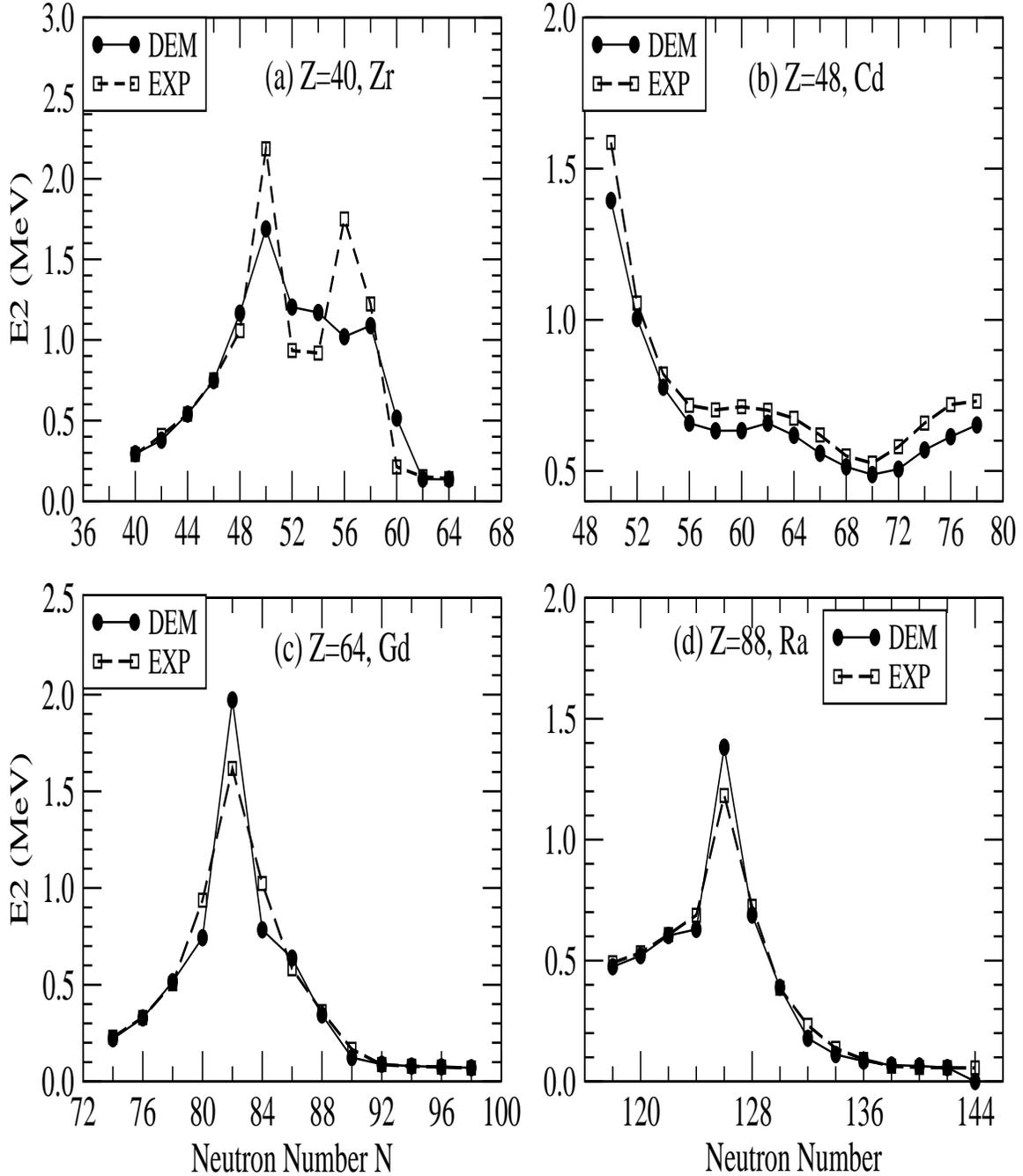}
\renewcommand{\baselinestretch}{1.}
\caption{ Similar to Fig. 5 but for Z=40, 48, 64 and 88. The experimental
data points marked as EXP
 correspond to those of Raman et al. \cite{rmn}.} 
\end{figure}
	We have just demonstrated as shown above the utility of the recursion
 relations for predicting E2 values for  some of the even-even isotopes
 of Cr, Fe, Ni and Zn in agreement with the latest experimental \cite{prt} data. Therefore it is desirable to find out whether
the model is   good enough   for such predictions in the higher mass regions of
 the nuclear chart. However in these regions there is no new data to compare
with and hence we can only compare with the  data  set of Raman et al. \cite{rmn}.
With this view,  we repeated our calculations 
 for higher isotope series 
following the same methodology outlined above. 
 Since the main aim of our
 present investigation is just to  establish the goodness of our model, we
 present  
 here results of only  few such series for which experimental data exist
for a relatively large number of isotopes. Accordingly  we have chosen
four isotope  series  Z=40, 48, 64 and 88 covering 
  nuclei both in mid-mass and heavy-mass regions. Our choice of the first
two  series namely Z=40 and 48 
 is again to see to what extent our model works across the semi-magic number
40 and magic number 50.  Our predictions  along with 
 those of experimental data of     Raman et al. \cite{rmn} are presented  
in Fig. 6. From the presented results we see that the 
agreement of the model values 
  with experiment is remarkably  good. All isotopic  variations of our model
predictions clearly follow those of the experiment. 
However small discrepancies exist at the magic  numbers N=50, 82
 and 126 as the DEM model need not be expected to hold good.

Now taking stock of all the results discussed so far,  one can fairly say that
 the
recursion relations for E2 work exceedingly   well almost throughout the nuclear chart.
 Even across the
magic numbers and sharply changing deformations, these relations have succeeded
in reproducing the experimental data to a large extent with a little bit
of deviation here and there. 
 In a nutshell, the recursion relations 
for the excitation energies  E2 derived here 
can be termed   sound enough as to have passed  the numerical test both in 
reproducing and predicting  the experimental data
,  and thereby establish  the goodness of the differential equation 
(\ref{be2}) from which they originate.
  
	\section{ Concluding Remarks}

In conclusion,  we  would like to say that  we have  succeeded in obtaining
 for the first time, a novel relation 
for the first $0_1 ^+ \rightarrow 2_1 ^+ $ state excitation energy E2 
 of a given even-even nucleus in terms of
its derivatives with respect to neutron and proton numbers.
We could establish  such a  differential equation  on the basis of  one-to-one 
correspondence with the local energy of the Infinite Nuclear Matter model
of atomic nuclei and 
the recently developed Differential Equation Model for
the complementary physical quantity B(E2)$\uparrow$.
   We have also succeeded in 
establishing its theoretical foundation on the basis of the empirical fact that
 E2's
are more or less slowly varying functions of neutron and proton numbers except
across the magic numbers.
We  further used the standard   definitions of the  derivatives with respect
to neutron and proton numbers  occurring in
 the equation,  to derive two recursion relations in E2. Both these relations 
are found to connect 
three different neighboring even-even nuclei from lower to higher mass  and
vice-verse.
The  numerical validity of these two relations  was further  established using
 the known experimental data set compiled by Raman et al. \cite{rmn} throughout
  the mass range of A=10 to 240.  
More importantly
 their utility was further demonstrated  by comparing our predictions 
with the latest experimental data set of Pritychenko et al. \cite{prt} for the
 isotopes of Cr, Fe, Ni and Zn.
The results so obtained convincingly show
the goodness of the recursion relations in E2 and thereby their parent differential
equation. 

{99}
\end{document}